\documentclass[
showpacs,
floatfix,
aps,
prl,
twocolumn,
superscriptaddress,
amssymb,
]{revtex4-1}

\usepackage{amssymb}
\usepackage{latexsym}
\usepackage[dvips]{graphicx}
\usepackage{amsmath}

\usepackage{graphicx}
\usepackage{dcolumn}
\usepackage{amsfonts}
\usepackage{bm}
\usepackage{epsfig}

\newcommand{\be}{\begin{equation}}
\newcommand{\ee}{\end{equation}}
\newcommand{\bea}{\begin{eqnarray}}
\newcommand{\eea}{\end{eqnarray}}

\def\oc{\omega_{\mbox{\scriptsize {c}}}}
\def\op{\omega_{\mbox{\scriptsize {p}}}}
\def\omp{\omega_{\mbox{\scriptsize {mp}}}}
\def\fp{f_{\mbox{\scriptsize {p}}}}

\def\ttr{\tau}

\def\op{\Omega}
\def\omp{\omega}
\def\oph{\omega_{\rm ph}}
\def\fp{F}

\newcommand{\req}[1]{Eq.\,(\ref{#1})}

\newcommand{\rfig}[1]{Fig.\,\ref{#1}}

\newcommand{\rref}[1]{Ref.\,\onlinecite{#1}}
\newcommand{\rrefs}[2]{Refs.\,\onlinecite{#1},\,\onlinecite{#2}}

\begin{document}
\title{Resistively detected high-order magnetoplasmons in a high-quality 2D electron gas}
\author{Q.~Shi}
\affiliation{School of Physics and Astronomy, University of Minnesota, Minneapolis, Minnesota 55455, USA}
\author{M.~A.~Zudov}
\email[Corresponding author: ]{zudov@physics.umn.edu}
\affiliation{School of Physics and Astronomy, University of Minnesota, Minneapolis, Minnesota 55455, USA}
\author{L.\,N. Pfeiffer}
\affiliation{Princeton University, Department of Electrical Engineering, Princeton, New Jersey 08544, USA}
\author{K.\,W. West}
\affiliation{Princeton University, Department of Electrical Engineering, Princeton, New Jersey 08544, USA}
\author{J.\,D. Watson}
\affiliation{Department of Physics and Astronomy, Purdue University, West Lafayette, Indiana 47907, USA}
\affiliation{Birck Nanotechnology Center, Purdue University, West Lafayette, Indiana 47907, USA}
\author{M.\,J. Manfra}
\affiliation{Department of Physics and Astronomy, Purdue University, West Lafayette, Indiana 47907, USA}
\affiliation{Birck Nanotechnology Center, Purdue University, West Lafayette, Indiana 47907, USA}
\affiliation{School of Materials Engineering, Purdue University, West Lafayette, Indiana 47907, USA}
 \affiliation{School of Electrical and Computer Engineering, Purdue University, West Lafayette, Indiana 47907, USA}

\begin{abstract}
We report on high-order magnetoplasmon resonances detected in photoresistance in high-mobility GaAs quantum wells. 
These resonances manifest themselves as a series of photoresistance extrema in the regime of Shubnikov-de Haas oscillations.
Extending to orders above 20, the extrema exhibit alternating strength, being less (more) pronounced at even (odd) order magnetoplasmon modes.
This experimental technique provides sensitive and elegant means to detect and investigate multiple magnetoplasmon modes and could be applied to other systems.
\end{abstract}
\pacs{73.43.Qt, 73.63.Hs, 73.40.-c}
\maketitle

When a two-dimensional electron gas (2DEG), laterally confined to a Hall bar of width $w$, is subjected to a weak perpendicular magnetic field $B$ and microwave radiation, its photoresistance reveals magneto-plasmon resonances (MPR) \citep{vasiliadou:1993,zudov:2001a,holland:2004,kukushkin:2006b,yang:2006,dorozhkin:2007b,tung:2009,andreev:2011,hatke:2012b,hatke:2013,note:2}.
Although there exists no quantitative theory of the MPR photoresistance, it is believed that the radiation absorption translates to electron heating which, in turn, causes a resistivity change \citep{vasiliadou:1993,kukushkin:2006b,dorozhkin:2007b}. 

The majority of photoresistance measurements revealed only the fundamental MPR, although in several cases the lowest few MPR modes were observed.
The $n$-th plasmon mode has a wavenumber $q_n = \pi n/w$ ($n = 1,2,...$) and due to the hybridization with the cyclotron mode in a magnetic field its dispersion takes the form  \citep{stern:1967,chaplik:1972}:
\be
\omp_n = \sqrt{\oc^2 + \op^2_n}\,,~~\op^2_n = n\op^2_1\,,~~ \op_1^2 = \frac{2\pi \alpha c}{\epsilon^\star w} \frac{E_F}{\hbar}\,.
\label{eq.1}
\ee
Here, $\oc = eB/m^\star$ is the cyclotron frequency of an electron with the effective mass $m^\star$, $\op_1 = 2\pi \fp_1$ is the frequency of the plasmon mode with the wavenumber $\pi/w$, $\alpha = e^2/4\pi\epsilon_0 \hbar c \approx 1/137$ is the fine structure constant, $\epsilon_0$ is the vacuum permittivity, 
$\epsilon^\star = (\epsilon+1)/2$ is the effective dielectric constant of the surrounding medium, $\epsilon=12.8$ is the dielectric constant of GaAs, and $E_F$ is the Fermi energy.

The MPR can occur when the excitation frequency $\omega = 2\pi f$ is equal to $\omp_n$ which can be tuned by $B$. 
Since the number of observable MPR modes is given by $(f/\fp_1)^2$,
 high excitation frequency $f$ is a prerequisite for observation of high-order modes.
In addition, for the MPR modes to be experimentally resolved, the spacing between them should exceed their width. 
The half-width of the $n$-th MPR can be written as \citep{chiu:1976,mikhailov:1996,zhang:2014}
\be
\delta\omp_n = \gamma + \Gamma_n\,,
\label{eq.2}
\ee
where $\gamma$ is the incoherent scattering rate and $\Gamma_n$ is the superradiant rate due to coherent dipole reradiation of oscillating 2D electrons.
The superradiant decay rate $\Gamma_n$ is given by \cite{leavitt:1986}
\be
\Gamma_n = \Gamma_0 (L_n/\lambda)^2\,,~~ \Gamma_0 = \frac {2\alpha}{n^\star}\frac{E_F}{\hbar}\,,
\label{eq.3}
\ee
where $L_n$ is the coherence length, $\lambda = c/f n^\star$ is the photon wavelength, and $n^\star = (\sqrt{\epsilon}+1)/2$ is the effective refractive index.
At high $n$, the superradiant rate drops rapidly due to the decay of coherence length $L_n \sim w/n$ \cite{muravev:2016} and the MPR width is  governed by the incoherent rate $\gamma$. 
Therefore, small incoherent scattering rate is another essential ingredient for the detection of high-order MPR.

In this Rapid Communication we report on high-order magnetoplasmon resonances resistively detected in very high-quality GaAs quantum wells at radiation frequencies up to $\sim 0.4$ THz.
Owing to high sensitivity of Shubnikov-de Haas oscillations (SdHO) to MPR-induced changes of electron temperature, the resonances manifest themselves as a series of resistance extrema superimposed on the SdHO.
Extending to orders as high as $n = 25$, the extrema exhibit alternating strength, being less (more) pronounced at even (odd) magnetoplasmon modes.
The employed experimental scheme provides a very efficient way to investigate the properties of MPR and could be applicable to other 2D systems. 

Our samples are 200 $\mu$m-wide Hall bars fabricated from symmetrically doped, 30-nm wide GaAs/AlGaAs quantum wells.
After low-temperature illumination, electron density and mobility of a Princeton-grown sample A were $n_e \approx 2.5 \times 10^{11}$ cm$^{-2}$ and $\mu \approx2.5 \times 10^7$ cm$^2$/Vs, respectively.
Sample B, grown at Purdue, had $n_e \approx 2.6 \times 10^{11}$ cm$^{-2}$ and $\mu \approx 1.3 \times 10^7$ cm$^2$/Vs.
The resistivity was measured at temperatures of $0.3-0.4$ K using a standard low-frequency lock-in technique, in sweeping $B$ and under radiation of frequency $f$ from 0.23 THz to 0.38 THz generated by a backward wave oscillator.

%%%%%%%%%%%%%%%%%%%%%%%
\begin{figure}[t]
%%\vspace{-0.2 in}
\includegraphics{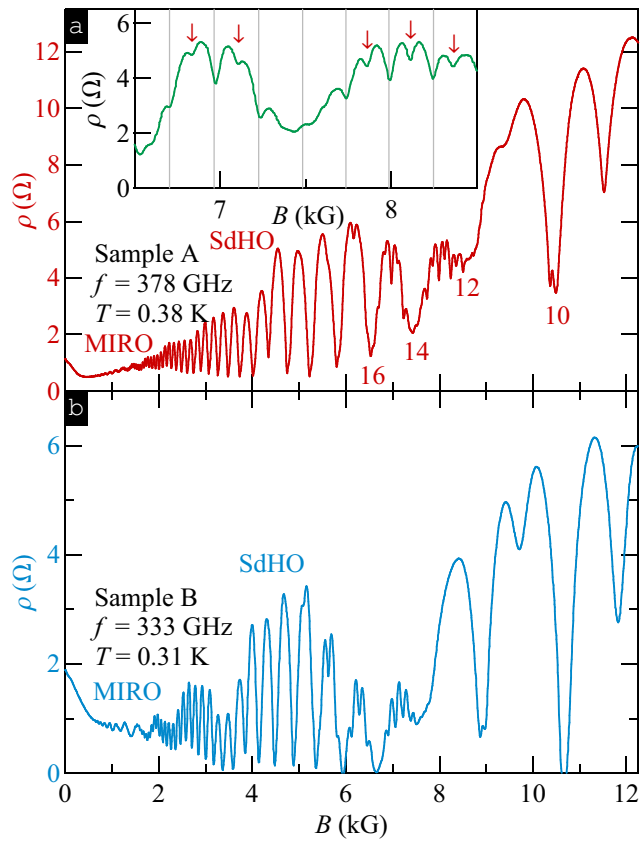}
%\vspace{-0.25 in}
\caption{(Color online)
(a) [(b)] $\rho (B)$ measured in sample A [B] at $f$ = 378 [333] GHz and $T \approx$ 0.38 [0.31] K. 
SdHO minima in (a) are marked by filling factor $\nu$ = 8, 10, 12, 14, 16. 
Inset shows zoomed-in part of the data shown in (a) near the cyclotron resonance. 
A series of deep (shallow) radiation-induced minima is marked by vertical lines (arrows).
}
%%\vspace{-0.2 in}
\label{fig1}
\end{figure}
%%%%%%%%%%%%%%%%%%%%%%%

In \rfig{fig1}(a)[(b))] we show magnetoresistance $\rho (B)$ under microwave irradiation of frequency $f$ = 378 GHz measured in sample A [B] at $f$ = 378 [333] GHz and $T \approx$ 0.38 [0.31] K.
In both samples, the data reveal microwave-induced resistance oscillations (MIRO) \citep{zudov:2001a,mani:2002,zudov:2003,yang:2003} and SdHO coexisting over a wide magnetic field range \cite{shi:2015b}.
In addition to MIRO and SdHO, both data sets show a structure consisting of multiple sharp features in the vicinity of the cyclotron resonance.

A closer look at this structure, shown in the inset of \rfig{fig1}(a), reveals that it is a series of extrema.
Near the SdHO maxima, this series is represented by resistance minima which are roughly equally spaced in the magnetic field;
between filling factors $\nu=12$ and $\nu = 14$, we observe at least seven such minima.
The minima can be divided into two classes, strong ones (cf. vertical lines) and weak ones (cf. $\downarrow$), appearing roughly in the middle between the neighboring strong minima.
As discussed below, these strong (weak) features originate from magnetoplasmon resonances of odd (even) order $n$, with $n$ as high as 25.
While the feature around $\nu = 10$ in \rfig{fig1}(a) also originates from photoresistance, identifying its exact nature is left for future studies. 
One possible origin is the plasmon-photon mode hybritization which leads to crossing of magnetoplasmon dispersion with cyclotron lines \citep{kukushkin:2003,muravev:2016}. 

Around the SdHO minima, MPRs are manifested by local maxima (see, e.g. a maximum at $B \approx 7.5$ GHz in the inset of \rfig{fig1}(a)). 
These maxima are less pronounced due to a weaker temperature dependence of the SdHO resistance minima in the regime of separated Landau levels. 
Indeed, at even $\nu$ the density of electron states is small in a wide band around the Fermi level and raising the temperature has little effect as long as it remains small compared to inter-level spacing. 

%%%%%%%%%%%%%%%%%%%%%%%
\begin{figure}[t]
%%\vspace{-0.2 in}
\includegraphics{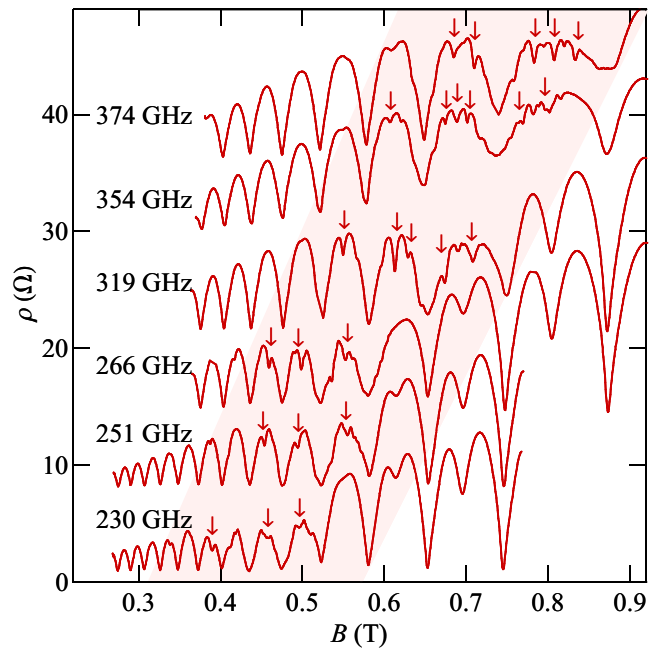}
%\vspace{-0.25 in}
\caption{(Color online)
$\rho (B)$ at different $f$, as marked, measured in sample A.
Some examples of MPR are marked by arrows.
The traces are offset for clarity by 7 $\Omega$.
}
%\vspace{-0.2 in}
\label{fig2}
\end{figure}
%%%%%%%%%%%%%%%%%%%%%%%

In \rfig{fig2} we plot magnetoresistance $\rho (B)$ measured in sample A at several frequencies from 230 GHz to 374 GHz, as marked \cite{note:6}.
Each trace reveals a series of extrema (some examples marked by arrow), similar to that shown in \rfig{fig1}, which move to higher $B$ with increasing $f$.
Since the resonances are most pronounced around the SdHO maxima, these extrema are best observed at higher $f$ (lower $\nu$) where more MPR modes can be resolved within one SdHO period.
One remarkable feature of our data is the unusually large number of both odd and even MPR modes, made possible by high sensitivity of SdHO to variations of electron temperature due to resonant absorption in our high mobility samples \cite{note:4}.

%%%%%%%%%%%%%%%%%%%%%%%
\begin{figure}[t]
%%\vspace{-0.2 in}
\includegraphics{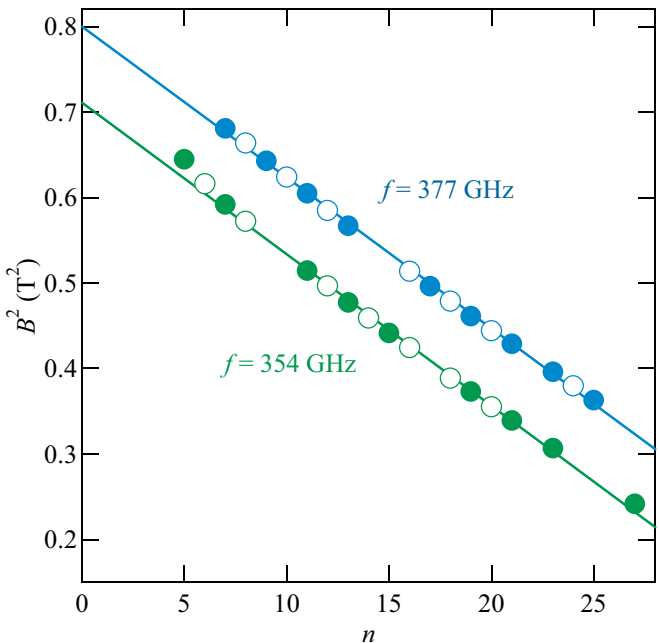}
%\vspace{-0.25 in}
\caption{(Color online)
$B^2$ (circles) vs. $n$, for $f = 377$ GHz (upper) and 354 GHz (lower), obtained from sample A.
Solid (open) circles represent odd (even) $n$.
Fits to the data (lines) using \req{eq.b} yield $m^\star \approx 0.066 m_0$ and $\fp \approx 56$ GHz.
}
%\vspace{-0.2 in}
\label{fig3}
\end{figure}
%%%%%%%%%%%%%%%%%%%%%%%

To confirm that the observed extrema originate from MPR we examine the dispersion relation for two radiation frequencies using the data from sample A.
Per \req{eq.1}, the magnetic field at which the $n$-th MPR ($\omega = \omp_n$, $B = B_n$) occurs should satisfy
\be
B^2 = B_n^2 \equiv \left ( m^\star/e \right )^2 \left [ \omega^2 - n\op_1^2 \right ]\,.
\label{eq.b}
\ee
In \rfig{fig3} we plot the square of the magnetic field, at which the $n$-th MPR occurs, for $f = 377$ GHz and $f = 354$ GHz, as marked, as a function of $n$. 
Here, filled (open) circles represent odd (even) $n$, spanning from $n = 5$ to $n =27$ \citep{note:1}.
Despite slight deviation at low $n$, the data for both frequencies conform to \req{eq.b}; linear fits generate $m^\star \approx 0.066 m_0$ \cite{note:3} and $\fp_1 \approx 56$ GHz.
Obtained $\fp_1$ is about 0.95 of what is expected from \req{eq.1}, in reasonable agreement with past theoretical \citep{mikhailov:2005} and experimental \citep{vasiliadou:1993,kukushkin:2006b,hatke:2012b} studies.

%%%%%%%%%%%%%%%%%%%%%%%
\begin{figure}[t]
%\vspace{0.2 in}
\includegraphics{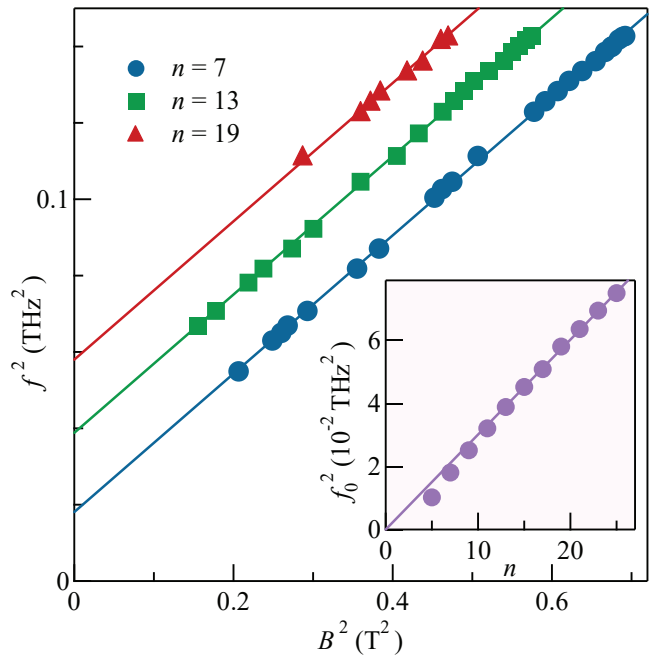}
%\vspace{-0.2 in}
\caption{(Color online)
$f^2$ vs. $B^2$ for $n = 7$ (circles), 13 (squares), 19 (triangles), obtained from sample A.
The fits to the data (lines) with $f^2 = f_0^2 + (e/2\pi m^\star)^2B^2$ yield $m^\star = 0.066m_0$.
Inset shows $f_0^2$, obtained from the fits, vs. $n$, revealing linear relationship, $f_0^2 \sim n$.
The slope of the linear fit (line) yields $\fp_1 \approx 55$ GHz.
}
%%\vspace{-0.2 in}
\label{fig4}
\end{figure}
%%%%%%%%%%%%%%%%%%%%%%%%

We next examine the frequency dependence of the observed resonances and demonstrate that it can also be well described by the magnetoplasmon dispersion, \req{eq.1}.
In \rfig{fig4} we plot $f^2$ vs. $B^2$ for three select modes, $n = 7, 13, 19$, and observe that the data can be well described by parallel lines.
Such a behavior is indeed expected for MPR since
\be
\omega^2 = \omp_n^2 \equiv n\op_1^2 + (e/m^\star)^2 B^2\,,
\ee
see \req{eq.1}.
The universal slope of the linear fits again conforms to $m^\star \approx 0.066 m_0$.

After performing such fits for all available $n$, we extract the intercepts of the fits with the vertical axis $f_0^2$ and present the result as a function of $n$ in the inset of \rfig{fig4}.
The data reveal a linear relation between $f_0^2$ and $n$, in accordance with \req{eq.1}, and from the slope of the fit we find $\fp_1 \approx 56$ GHz, in good agreement with the value obtained above.
We further note that at $n < 10$, $f_0^2 (n)$ slightly deviates from the linear dependence and bends down.
As we show next, this deviation is likely due to retardation effects.

The retardation effects are characterized by the retardation parameter $\alpha^{\prime}$, defined as the ratio of the plasmon frequency to the light frequency at the same wavenumber.
At the fundamental plasmon wavenumber $q_1 = \pi/w$, we find, for sample A, $\oph/2\pi = c/2w n^\star \approx 0.33$ THz and $\alpha' = \Omega_1/\oph \approx 0.07(E_F/\hbar\oph)^{1/2} \approx 0.18$.
Since the phase velocity of the magnetoplasmon mode is proportional to $n^{-1/2}$, the retardation effects become weaker at higher $n$.
It has been shown that the plasmon frequency decreases with $\alpha^{\prime}$ \cite{kukushkin:2003,kukushkin:2006b}. 
It is therefore reasonable to have a larger deviation from \req{eq.1} at lower $n$, as observed in both \rfig{fig3} and \rfig{fig4}(b).

As mentioned in the introduction, observation of multiple MPR modes relies on their sharpness. 
We estimate the half-width of the observed MPR as $\delta B_n \approx$ 2.5 mT which is largely insensitive to $n$.
Since the width is obtained from magnetoresistance, it does not directly translate to the width of the MPR.
Nevertheless, since the width of the photoresistance minima does not show significant variation within one SdHO period, it should be comparable to the actual MPR width.
Using $\delta\omp_n \approx (\partial \omp_n/\partial B) \delta B_n = (e/m^\star)(\oc/\omega)\delta B_n$, we estimate $\delta\omp_n/2\pi \simeq 1$ GHz, close to the cyclotron resonance.

The MPR width $\delta\omega_n$ is expected to be a sum of coherent radiative decay rate $\Gamma_n$ and incoherent collisional rate $\gamma$, see \req{eq.2} and \req{eq.3}.
In sample A we estimate $\Gamma_0/2\pi \approx 14$ GHz and at $f = 0.38$ GHz, $\Gamma_1/2\pi = \Gamma_0/2\pi (w/\lambda)^2 \approx 4.7$ GHz.
At higher $n$, $\Gamma_n = \Gamma_1/n^2$ becomes smaller than $\gamma$.
This is consistent with our observation that the fundamental MPR $n = 1$ is not well resolved, whereas higher order MPRs have similar widths, dominated by $\gamma$.

The incoherent scattering rate $\gamma$ is often associated with the transport scattering rate $\ttr^{-1}$, which is about $\ttr^{-1}/2\pi \approx 0.2$ GHz in sample A. 
However, there exist no exact relation between the two quantities; for the electron density of sample A, \rref{andreev:2014} has found $\gamma$ to be about 5 times larger than $\ttr^{-1}$, giving a conservative estimate of $\gamma/2\pi \simeq$ 1 GHz.
This value is in good agreement with $\delta\omega_n/2\pi \simeq$ 1 GHz obtained from our experimental data.

%%%%%%%%%%%%%%%%%%%%%%%
\begin{figure}[t]
%%\vspace{-0.2 in}
\includegraphics{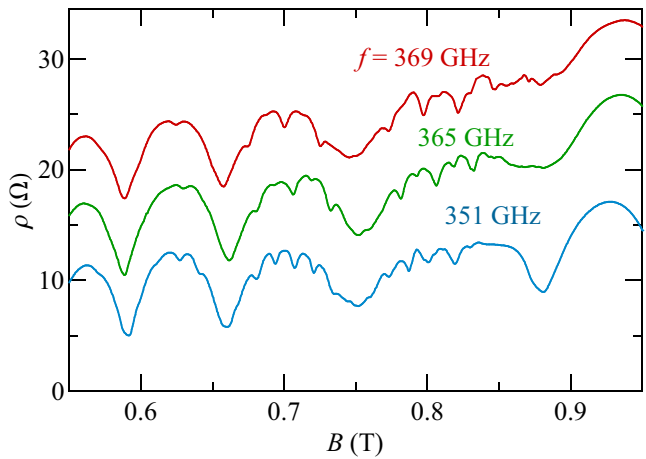}
%\vspace{-0.2 in}
\caption{(Color online)
$\rho (B)$ at $T \approx$ 0.4 K and different $f$, as marked, measured in sample A.
Traces are offset for clarity by 7 $\Omega$.
}
%\vspace{-0.2 in}
\label{fig5}
\end{figure}
%%%%%%%%%%%%%%%%%%%%%%%

Another interesting feature revealed by our data is the even-odd mode alternating strength \citep{note:8}.
Assuming a uniform electric field distribution, only odd modes are expected \cite{mikhailov:2005} and \rrefs{kukushkin:2006b}{bialek:2014} \cite{note:7} reported observation of only odd modes.
Some experiments, however, revealed both odd and even low-order modes, with often comparable, yet random strengths \cite{studenikin:2007,baskin:2011}.
While we have clearly observed different signal strengths for even and odd modes, their ratio can vary with the radiation frequency.
In \rfig{fig5} we present $\rho(B)$ measured in sample A at different $f$, 369 GHz, 365 GHz, and 351 GHz.
At $f=369$ GHz (top trace), the even modes are much weaker than the odd, at $f = 365$ GHz (middle trace) they become noticeably stronger, and at $f=351$ GHz (bottom trace) even modes become comparable to the odd ones. 
The variation in the strength of even modes with $f$ likely reflects variation in radiation distribution \cite{note:5}.
While a more uniform radiation produces weaker even modes, they become more pronounced under nonuniform radiation distribution. 
Our measurements never revealed even modes which are stronger than the odd ones.

In summary, exploiting high mobility 2DEGs, high radiation frequencies, and strong sensitivity to temperature of SdHO, we have detected high-order magnetoplasmon resonances. 
Extending to orders above $n = 25$, the MPR-induced photoresistance extrema exhibit alternating strength, being less (more) pronounced at even (odd) $n$.
Taken together, our results demonstrate that this experimental technique provides a surprisingly sensitive and elegant means to detect and investigate high-order MPR modes which could be applicable to other 2D systems, such as MgZnO/ZnO heterojunctions \citep{kozlov:2014} and AlAs quantum wells \citep{muravev:2015}.

\begin{acknowledgments}
We thank I. Kukushkin and V. Muravev for discussions and G. Jones, S. Hannas, A. Hatke, T. Murphy, J. Park, D. Semenov, D. Smirnov and K. Thirunavukkuarasu for technical assistance. 
The work at Minnesota (Purdue) was supported by the U.S. Department of Energy, Office of Science, Basic Energy Sciences, under Award \# ER 46640-SC0002567 (DE-SC0006671).
The work at Princeton University was funded by the Gordon and Betty Moore Foundation through the EPiQS initiative Grant GBMF4420, and by the National Science Foundation MRSEC Grant DMR-1420541.
Q.S. acknowledges University of Minnesota Doctoral Dissertation fellowship.
Experiments were performed at the National High Magnetic Field Laboratory, which is supported by NSF Cooperative Agreement No. DMR-0654118, by the State of Florida, and by the DOE.
\end{acknowledgments}

%\bibliographystyle{apsrev}
%\bibliography{bibRMP1qs,footnotes}

%\bibliographystyle{../../apsrev-titles}
%\bibliographystyle{../../apsrev}
%\bibliography{../../bibRMP1qs.1_2,footnotes}

\begin{thebibliography}{37}
\bibitem[{not({\natexlab{b}})}]{note:6}
\bibinfo{note}{While at different $f$ we sometimes detect variations of
  electron density of $\sim 1\%$ due to different power, these variations have
  negligible effect on the MPR dispersion.}
\bibitem[{not({\natexlab{h}})}]{note:5}
\bibinfo{note}{Our experiments are performed without using resonant cavities
  and, as a result, the radiation field distribution is not controllable.
  Furthermore, the presence of contacts, metallic pins on the chip carrier, and
  other metallic parts make random formation of standing waves in the sample
  space possible. At 0.4 THz, the distance between the nodes of such standing
  waves is close to or smaller than the sample dimensions. As a result,
  different parts of the sample are exposed to different radiation
  intensities.}
\end{thebibliography}

\end{document}